\documentclass[prd,aps,nofootinbib,preprintnumbers,preprint]{revtex4}
\usepackage{graphicx}

\begin{document}

\newcommand{\gsim}{ \mathop{}_{\textstyle \sim}^{\textstyle >} }
\newcommand{\lsim}{ \mathop{}_{\textstyle \sim}^{\textstyle <} }
\newcommand{\vev}[1]{ \left\langle {#1} \right\rangle }

\newcommand{\bear}{\begin{array}}  \newcommand{\eear}{\end{array}}
\newcommand{\bea}{\begin{eqnarray}}  \newcommand{\eea}{\end{eqnarray}}
\newcommand{\beq}{\begin{equation}}  \newcommand{\eeq}{\end{equation}}
\newcommand{\bef}{\begin{figure}}  \newcommand{\eef}{\end{figure}}
\newcommand{\bec}{\begin{center}}  \newcommand{\eec}{\end{center}}
\newcommand{\non}{\nonumber}  \newcommand{\eqn}[1]{\beq {#1}\eeq}
\newcommand{\la}{\left\langle} \newcommand{\ra}{\right\rangle}

\def\lrf#1#2{ \left(\frac{#1}{#2}\right)}
\def\lrfp#1#2#3{ \left(\frac{#1}{#2}\right)^{#3}}

%

\renewcommand{\thefootnote}{\alph{footnote}}

\renewcommand{\thefootnote}{\fnsymbol{footnote}}
\preprint{DESY 06-035}
\title{Moduli/Inflaton Mixing with Supersymmetry Breaking Field}
\renewcommand{\thefootnote}{\alph{footnote}}

\author{Motoi Endo$^{1,2}$, Koichi Hamaguchi$^{1,3}$ and Fuminobu Takahashi$^{1,2}$}

\affiliation{
${}^1$ Deutsches Elektronen Synchrotron DESY, Notkestrasse 85,
  22607 Hamburg, Germany\\
${}^2$ Institute for Cosmic Ray Research,
  University of Tokyo, Chiba 277-8582, Japan\\
${}^3$ Department of Physics, University of Tokyo, Tokyo 113-0033, Japan
  }

\begin{abstract}
\noindent
A heavy scalar field such as moduli or an inflaton generally mixes
with a field responsible for the supersymmetry breaking.  We study the
scalar decay into the standard model particles and their
superpartners, gravitinos, and the supersymmetry breaking sector,
particularly paying attention to decay modes that proceed via the
mixing between the scalar and the supersymmetry breaking field.  The
impacts of the new decay processes on cosmological scenarios are also
discussed; the modulus field generically produces too many gravitinos,
and most of the inflation models tend to result in too high reheating
temperature and/or gravitino overproduction.
\end{abstract}

\maketitle

\section{Introduction}
Scalar fields play an important role in the thermal history of the
universe.  Once a scalar field dominates the energy density of the
universe, the subsequent evolution of the universe strongly depends on
the reheating processes characterized by the decay temperature and the
decay products.

Such scalar fields, symbolically denoted by $\phi$, may be identified
with an inflaton or moduli fields.  A modulus field generally acquires
nonzero vacuum expectation value (VEV) in the vacuum. Inflaton fields
as well have non-vanishing VEVs in many inflation models. Once a
scalar field obtains a nonzero VEV, $\phi_0 \equiv \la\phi\ra$, there
is no remnant symmetry to forbid mixings of $\phi$ with the other
fields, since the symmetries under which $\phi$ is charged, if any,
are spontaneously broken in the vacuum.  There is another important
scalar field, $z$, which is responsible for the supersymmetry (SUSY)
breaking. The presence of such SUSY breaking field is inevitable in
the SUSY theories, because of an absence of the light
superparticles. The SUSY breaking field, $z$, must be singlet under
any unbroken symmetries at the vacuum in order for the auxiliary
field, $G_z$, to obtain a finite VEV. Therefore the scalar field $z$
as well generally obtains a VEV, $z_0 \equiv \la z \ra$.

We would like to stress that a scalar field $\phi$ with nonzero VEV,
such as the inflaton and moduli, generically {\it mixes} with the SUSY
breaking field $z$ in the vacuum.  In particular such mixing has
impacts on the decay processes of $\phi$.  It has been recently argued
that the modulus and inflaton decays may produce too many gravitinos
and/or the lightest SUSY particle
(LSP)~\cite{Endo:2006zj,Nakamura:2006uc,Kawasaki:2006gs,Asaka:2006bv}.
In Ref.~\cite{Dine:2006ii}, however, it has been demonstrated that the
gravitino production rate can be suppressed by taking account of the
mixing of $\phi$ with $z$ in some explicit models. In this paper, we
develop general analyses on the mixture of $\phi$ and $z$, and discuss
its cosmological consequences, paying particular attention to the
decay of $\phi$ via the mixing with $z$.

In the next section, we develop a formalism to obtain the
mass-eigenstate basis and clarify the relation between the
mass-eigenstate basis and the model basis.  In Sec.~\ref{sec:3}, we
consider several decay processes in the mass eigenstates, especially
those induced via the mixing with the SUSY breaking sector, in the
gravity-mediated SUSY breaking scenario.  We also discuss how the
modulus and inflaton cosmology is affected by the mixing. In
Sec.~\ref{sec:4} we take up the low energy SUSY breaking models such
as the gauge-mediated SUSY breaking (GMSB) models~\cite{GMSB},
clarifying the difference from the case of gravity mediation.  
Sec.~\ref{sec:5} is devoted to discussions on miscellaneous topics.
We give a summary in the last section.
In Appendix.~\ref{appa}, we show the goldstino interpretation of the
scalar decay into gravitinos and see the equivalence between the two pictures.

\section{Mass-eigenstate Basis}
\label{sec:2}
A scalar decay must be considered in its mass-eigenstate basis, while
a model is often given in such a way that particles in the model are
not mass eigenstates especially if some symmetries are spontaneously
broken in the vacuum. In particular, it is quite probable that a
scalar $\phi$ with nonzero VEV mixes with the SUSY breaking field $z$
in the vacuum, since there is no remnant symmetry that forbids the
mixing.  The kinetic term and non-analytic (NA) and analytic (A) mass
terms of $\phi$ and $z$ in the model frame are given as
\begin{eqnarray}
  && {\mathcal L}_{\rm kin.} \;=\;
  \partial_\mu \phi^\dagger \partial^\mu \phi + 
  \partial_\mu z^\dagger \partial^\mu z + 
  g_{z \bar \phi} \partial_\mu \phi^\dagger \partial^\mu z + 
  g_{\phi \bar z } \partial_\mu z^\dagger \partial^\mu \phi, 
  \\
  && -{\mathcal L}_{\rm mass}^{\rm (NA)} \;=\;
  M_{\phi \bar \phi}^2 \phi^\dagger \phi + 
  M_{z \bar z}^2 z^\dagger z + 
  M_{z \bar \phi}^2 \phi^\dagger z + 
  M_{\phi \bar z}^2 z^\dagger \phi,
  \\
  && -{\mathcal L}_{\rm mass}^{\rm (A)} \;=\; 
  {1\over 2} M_{\phi \phi}^2 \phi \phi + 
  {1\over 2} M_{z z}^2 z z + 
  M_{\phi z}^2 \phi z + {\rm h.c.},
\end{eqnarray}
where the fields are expanded around the VEV, 
$\phi \rightarrow \phi - \phi_0$ and $z \rightarrow z - z_0$.
The mixings  in the kinetic term, $g_{z \bar \phi}$ 
and  $g_{\phi \bar z}$,  are given by
\bea
g_{z \bar \phi} &=& \la \frac{\partial^2 K}{\partial z \partial \phi^\dag} \ra,~~~
g_{\phi \bar z} = \la \frac{\partial^2 K}{\partial \phi  \partial z^\dag}\ra,
\eea
where $K$ is the K\"ahler potential, while $g_{\phi \bar \phi}$ and
$g_{z \bar z}$ are normalized to be unity.  Note that the cross term
$g_{z \bar \phi} \partial_\mu \phi^\dagger \partial^\mu z$ naturally
appears if there are higher order terms in the K\"ahler potential
before the fields are expanded around the VEV.  The purpose of this
section is to clarify the relation between the model basis $(\phi, z)$
and the mass-eigenstate basis.

In the Einstein frame, the 4D $N=1$ supergravity (SUGRA) Lagrangian
contains the scalar potential, $V = e^G (G^i G_i - 3)$~\footnote{
Throughout this paper we assume that the D-term potential is negligible.
}.  The
(non-)analytic mass terms can be written in terms of the total
K\"ahler potential, $G = K + \ln|W|^2$, as
\begin{eqnarray}
  M_{ij*}^2 &=& \frac{\partial^2 V}{\partial \varphi^i \partial \varphi^{\dagger j}}
  \;=\; e^G \left( 
    \nabla_i G_k \nabla_{j*} G^k - R_{ij^*k\ell^*} G^k G^{\ell*} + g_{ij^*}
  \right),
  \label{eq:Mijs}
  \\
  M_{ij}^2 &=& M_{ji}^2 
  \;=\; \frac{\partial^2 V}{\partial \varphi^i \partial \varphi^j}
  \;=\; e^G \left( 
    \nabla_i G_j + \nabla_j G_i + G^k \nabla_i \nabla_j G_k 
  \right),
  \label{eq:Mij}
\end{eqnarray}
where we have assumed the vanishing cosmological constant, $G^i G_i =
3$, and used the potential minimization condition, $G^i \nabla_k G_i +
G_k = 0$ in the vacuum. The gravitino mass is given by $m_{3/2} = \la
e^{G/2} \ra$.  Here and in what follows, the subscript $i$ denotes a
derivative with respect to the field $\varphi^i$, and the superscript
is defined by $G^i = g^{ij^*} G_{j*}$. Here $g_{ij^*}$ is the K\"ahler
metric, $g_{ij^*} = G_{ij^*}$, and $R_{ij^*k\ell^*}$ is the curvature
of the K\"ahler manifold, defined by $R_{ij^*k\ell^*} =
g_{ij^*k\ell^*} - g^{mn^*} g_{mj^*\ell^*} g_{n^*ik}$.  Also the
covariant derivative of $G_i$ is defined by $\nabla_i G_j = G_{ij} -
\Gamma^k_{ij} G_k$, where the connection, $\Gamma_{ij}^k = g^{k\ell^*}
g_{ij\ell^*}$, and $\nabla_k g_{ij^*} = 0$ is satisfied.

Throughout this paper, the scalar field, $\phi$, is assumed to be much
heavier than the gravitino due to a large supersymmetric mass,
$m_\phi/m_{3/2} \equiv |\nabla_\phi G_\phi| \gg 1$.  The SUSY breaking
field $z$ is such that it sets the cosmological constant to be zero,
i.e., $G^z G_z \simeq 3$, while $\phi$ is assumed to give only
subdominant contribution to the SUSY breaking, i.e., $|G_\phi| \ll 1$.
Then, as long as $|\nabla_\phi G_z| \lesssim O(1)$~\footnote{
In fact, according to the discussion of Ref.~\cite{Endo:2006zj},
$|\nabla_\phi G_z|\sim O(1)$ holds for modulus field with its VEV of
the Planck scale, if the K\"ahler potential does not have any
enhancement factor.  In principle, $|\nabla_\phi G_z|$ could be larger
than O(1) if the higher order term $g_{\phi z \bar z}$ in the K\"ahler
potential is larger than unity.  However, such a large mixing in the
supersymmetric mass obscures the definitions (or roles) of the
different two fields in the model basis.  Also it makes the gravitino
problem even worse.
},
the potential minimization
condition for $z$,
\begin{eqnarray}
  G^z \nabla_z G_z + G^\phi \nabla_z G_\phi + G_z &=& 1\,,
\end{eqnarray}
requires that the supersymmetric mass of $z$ is equal to the gravitino
mass, i.e., $|\nabla_z G_z| \simeq 1$~\footnote{
This was also noted in Ref.~\cite{Lebedev:2006qq} in a different context.
}.  We assume that this is the
case.  It should be noted, however, that the scalar mass of $z$ can be
larger than $m_{3/2}$ due to the non-supersymmetric mass term, $e^G
R_{z \bar{z} k\ell^*} G^k G^{\ell^*}$, if one adds, e.g. $\delta K =
-|z|^4/\Lambda^2$ with a low cut-off scale $\Lambda \ll M_P$, to the
K\"ahler potential, which leads to $m_z^2\simeq 12 m_{3/2}^2 (M_P /
\Lambda)^2$.

In the following we assume $M^2_{\phi \bar \phi}$ dominates over the
other components of the mass terms.  The results in the case of
$M^2_{z \bar z} \gg M^2_{\phi \bar \phi} \gg ({\rm the~other~elements})$ will
be given in the last of this section. The rest discussion of this
section is however rather generic, and can be applied not only to the
situation we stated above.

The kinetic term can be canonically normalized by a shift of $z$ and a
rescaling of $\phi$;
\begin{eqnarray}
  \phi' &=& (1-|g_{\phi \bar z }|^2)^{-1/2} \phi,
  \\
  z' &=& z + g_{\phi \bar z } \phi ,
  \\
  {\mathcal L}_{\rm kin.} &=&
  \partial_\mu \phi'^\dagger \partial^\mu \phi' + 
  \partial_\mu z'^\dagger \partial^\mu z'.
\end{eqnarray}
Then the mass terms become
\begin{eqnarray}
  -{\mathcal L}_{\rm mass}^{(NA)} 
  &\equiv&   M_{\phi' \bar \phi'}^2  \phi'^\dagger \phi'
  + M_{z' \bar z'}^2 z'^\dagger z'+
  M_{z' \bar \phi'}^2 \phi'^\dagger z'
  +
  M_{\phi' \bar z'}^2 z'^\dagger \phi' \non\\
  &\simeq&
  (M_{\phi \bar \phi}^2 - g_{\phi \bar z}M_{z \bar \phi}^2 - g_{z \bar \phi}M_{\phi \bar z}^2)
  \phi'^\dagger \phi'
  +
  M_{z \bar z}^2 z'^\dagger z'
  \nonumber \\ &&+
  (M_{z \bar \phi}^2 - g_{z \bar \phi} M_{z\bar z}^2)\phi'^\dagger z'
  +
  (M_{\phi \bar z}^2 - g_{\phi \bar z} M_{z\bar z}^2)z'^\dagger \phi',
  \\
  -{\mathcal L}_{\rm mass}^{(A)} 
  &\simeq& 
  {1\over 2}(M_{\phi \phi}^2 - 2 g_{\phi \bar z} M_{\phi z}^2 ) \phi' \phi' + 
  {1\over 2}M_{z z}^2 z'z'
  \nonumber \\ &&+
  (M_{\phi z}^2 - g_{\phi \bar z}M_{zz}^2) \phi' z' + 
  {\rm h.c.}, 
\end{eqnarray}
where $\mathcal{O}(g_{\phi\bar z}^2)$ terms are omitted in each term.

Let us first diagonalize the non-analytic mass terms while keeping 
the kinetic term canonical by the following transformation:
\begin{eqnarray}
  \Phi &\equiv& \phi' + \epsilon z',\non\\
   Z &\equiv& z' - \epsilon^* \phi',
\end{eqnarray}
where $\epsilon$ represents the mixing angle. Here we have assumed
$|\epsilon| \ll 1$ and neglected those terms of $O(\epsilon^2)$.
Since $M_{\phi' \bar\phi'}^2$ dominates over the other components in
the mass matrix, the mixing angle is given by the ratio of $M_{\phi'
\bar\phi'}^2$ to the off-diagonal component:
\beq
\label{eq:epsilon}
\epsilon \simeq  \frac{M_{z' \bar \phi'}^2}{M_{\phi' \bar \phi'}^2} 
\simeq \frac{M_{z \bar \phi}^2 - g_{z \bar \phi} M_{z \bar z}^2}{M_{\phi \bar \phi}^2}.
\eeq
Then the non-analytic mass matrix becomes diagonal in this basis:
\begin{eqnarray}
  -{\mathcal L}_{\rm mass}^{(NA)} &\simeq&
 M_{\phi \bar \phi}^2  \Phi^\dagger \Phi
  +
  \left(M_{z \bar z}^2 - \frac{|M^2_{\phi \bar z}|^2}{M^2_{\phi \bar \phi}}\right) 
  Z^\dagger Z,
\end{eqnarray}
We will call this basis $(\Phi, Z)$ as the NA mass-eigenstate basis in
the following.

The physical processes become easy to be considered after the
mass matrix is fully diagonalized. In particular, one should note that
the analytic mass terms provide a further mixture between $\phi$ and
$z^\dag$ ($z$ and $\phi^\dag$). In the NA mass-eigenstate basis, the
analytic mass terms become
\begin{eqnarray}
  -{\mathcal L}_{\rm mass}^{(A)} &\equiv& 
  \frac{1}{2}M_{\Phi\Phi}^2 \Phi\Phi + \frac{1}{2}M_{ZZ}^2 ZZ + 
  M_{\Phi Z}^2 \Phi Z + {\rm h.c.}
  \nonumber \\
  &\simeq& 
  \frac{1}{2} \left( M_{\phi \phi}^2 - 2  ( g_{\phi \bar z } - \epsilon^* ) 
    M_{\phi z}^2 \right) \Phi \Phi + 
  \frac{1}{2}  \left( M_{z z}^2 - 2  \epsilon M_{\phi z}^2\right) ZZ 
  \nonumber \\ &&
  + \left( M_{\phi z}^2 - M_{zz}^2 g_{\phi \bar z } 
  -  M_{\phi\phi}^2 \epsilon+ M_{zz}^2 \epsilon^* \right) \Phi Z + {\rm h.c.}
\end{eqnarray}
up to $O(\epsilon)$. Let us concentrate on the mixings $\Phi - Z^\dag$ 
and $Z - \Phi^\dag$, since they
become important in the following analyses.  To diagonalize the
analytic mass term, we take the following transformation,
\begin{eqnarray}
\label{eq:a-phi}
  \tilde \Phi &\equiv& \Phi + \tilde \epsilon Z^\dagger, \\
\label{eq:a-z}
  \tilde Z    &\equiv& Z - \tilde \epsilon \Phi^\dagger,
\end{eqnarray}
where we have assumed that the mixing angle $\tilde \epsilon$ is much smaller
than unity.  Since the dominant contribution to the total mass matrix
comes from the non-analytic mass component, $M_{\Phi\bar\Phi}^2 \simeq
m_\phi^2$, $\tilde \epsilon$ is given by the ratio of $M_{\Phi\bar\Phi}^2$ to
$M_{\bar \Phi \bar Z}^2$:
\beq
\label{eq:tilep}
\tilde \epsilon\;\simeq\; \frac{M_{\bar \Phi \bar Z}^2}{M_{\Phi \bar\Phi}^2}
\simeq \frac{M_{\bar \phi \bar z}^2 - g_{z \bar \phi} M_{\bar z \bar z}^2}{M_{\phi \bar\phi}^2}.
\eeq
Thus obtained $(\tilde \Phi, \tilde Z)$ is the desired mass-eigenstate
basis.  Although $M_{\Phi\Phi}^2\,(M_{ZZ}^2)$ can further mix the real and
imaginary components of $\Phi\,(Z)$, it does not modify the following
discussions qualitatively.

Here, we summarize the relation between the model basis $(\phi, z)$
and the mass-eigenstate basis $(\tilde \Phi, \tilde Z)$.
\bea
\label{eq:rel}
\phi &\simeq& \tilde{\Phi} - \epsilon \tilde{Z} - \tilde \epsilon \tilde{Z}^\dag,\\
\label{eq:rel2}
z &\simeq& \tilde{Z} +(-g_{\phi \bar z} + \epsilon^*) \tilde{\Phi} + 
\tilde \epsilon \Phi^\dag.
\eea
The explicit expressions for $\epsilon$ and $\tilde \epsilon$ 
are given by (\ref{eq:epsilon}) and (\ref{eq:tilep}).

So far, we have assumed that $M^2_{\phi \bar \phi}$ dominates over the
other components of the mass matrix. If $z$ acquires a
non-supersymmetric mass larger than the mass of $\phi$, we can repeat
a similar discussion to obtain the relation between the model basis
and the mass-eigenstate basis:
\bea
\label{eq:rel3}
\phi &\simeq& \tilde{\Phi} +(-g_{z \bar \phi} + \epsilon^*) \tilde{Z} + 
\tilde \epsilon Z^\dag,\\
z &\simeq& \tilde{Z} - \epsilon \tilde{\Phi} - \tilde \epsilon \tilde{\Phi}^\dag
\label{eq:rel4}
\eea
with  $\epsilon$ and $\tilde \epsilon$ given by
\bea
\epsilon &\simeq&  \frac{M_{\phi' \bar z'}^2}{M_{z' \bar z'}^2} 
\simeq \frac{M_{\phi \bar z}^2 - g_{\phi \bar z} M_{\phi \bar \phi}^2}{M_{z \bar z}^2},\\
\tilde \epsilon &\simeq& \frac{M_{\bar Z \bar \Phi}^2}{M_{Z \bar Z}^2}
\simeq \frac{M_{\bar z \bar \phi}^2 - g_{\phi \bar z} M_{\bar \phi \bar \phi}^2}{M_{z \bar z}^2}.
\label{eq:with-c}
\eea

 In the following sections, we are particularly interested in the
 mixing of $\tilde \Phi (\tilde \Phi^\dag)$ into $z$ or $\tilde Z
 (\tilde Z^\dag)$ into $\phi$.  Although three sets of the
 transformations are necessary to arrive at the mass eigenstate
 basis, the effective mixing angle is given by the largest mixing
 among them.  To parametrize this effective mixing of $\tilde \Phi
 (\tilde \Phi^\dag)$ into $z$, we define $\epsilon_{z \tilde{\Phi}}$
 as
\beq
\epsilon_{z \tilde{\Phi}}\;\equiv \;\left\{
\bear{cc}
\displaystyle{ {\rm Max} \left\{|g_{\phi \bar z}|,\, 
    \left|\frac{M^2_{\phi \bar z}}{M^2_{\phi \bar \phi}}\right|,
\, \left|\frac{M^2_{\phi  z}}{M^2_{\phi \bar \phi}}\right| \right\} }
&~~~{\rm for}~~~M^2_{\phi \bar \phi} \gg M^2_{z \bar z},\\
&\\
\displaystyle{{\rm Max} \left\{\left|g_{\phi \bar z} \frac{M_{\phi \bar \phi}^2}{M_{z \bar z}^2} \right|,\,
 \left|\frac{M^2_{\phi \bar z}}{M^2_{z \bar z}}\right|,
\, \left|\frac{M^2_{\phi  z}}{M^2_{z \bar z}}\right| \right\}}
&~~~{\rm for}~~~M^2_{z \bar z} \gg M^2_{\phi \bar \phi}.
\eear
\right.
\eeq

Similarly, we define the effective mixing of $\tilde Z (\tilde
Z^\dag)$ into $\phi$ as
\beq
\epsilon_{\phi \tilde{Z}} \;\equiv \;\left\{
\bear{cc}
\displaystyle{ {\rm Max} \left\{\left|g_{\phi \bar z} \frac{M_{z \bar z}^2}{M_{\phi \bar \phi}^2} \right|,\,\left|
      \frac{M^2_{\phi \bar z}}{M^2_{\phi \bar \phi}}\right|,
\, \left|\frac{M^2_{\phi  z}}{M^2_{\phi \bar \phi}}\right| \right\} }
&~~~{\rm for}~~~M^2_{\phi \bar \phi} \gg M^2_{z \bar z},\\
&\\
\displaystyle{{\rm Max} \left\{|g_{\phi \bar z}|,\, 
    \left|\frac{M^2_{\phi \bar z}}{M^2_{z \bar z}}\right|,
\, \left|\frac{M^2_{\phi  z}}{M^2_{z \bar z}}\right| \right\}}
&~~~{\rm for}~~~M^2_{z \bar z} \gg M^2_{\phi \bar \phi}.
\eear
\right.
\eeq
Therefore, using these the effective mixing angles, the relations
(\ref{eq:rel}), (\ref{eq:rel2}), (\ref{eq:rel3}), and (\ref{eq:rel4}) 
can be roughly expressed as
\bea
\label{eq:rough-rel}
\phi &\sim& \tilde{\Phi} - \epsilon_{\phi \tilde{Z}} \tilde{Z}\\
\label{eq:rough-rel2}
z &\sim& \tilde{Z} +\epsilon_{z \tilde{\Phi}} \tilde{\Phi},
\eea
up to  phase, where we have also dropped the distinction 
between $\tilde \Phi (\tilde Z)$ and its conjugate.

\section{Gravity Mediation}
\label{sec:3}
Let us now consider the decay of $\tilde \Phi$ 
via the mixing with the SUSY breaking
field $z$, and discuss its cosmological influence.  
To this end, we need to specify the 
way to mediate the SUSY breaking to the visible sector. 
In this section we consider the
gravity mediation to exemplify how serious the problems 
caused by the mixing is.

\subsection{Decay Modes}

Let us study the scalar decay modes which proceed via the mixing with
the SUSY breaking field. They are classified by the decay products:
(i) the gravitinos; (ii) the SM particles (and their superpartners);
(iii) the SUSY breaking fields. We discuss each case below.

\subsubsection{Gravitino}

The scalar field $\phi$ can decay into a pair of the gravitinos
through the mixing with the SUSY breaking field~\footnote{
If the large scalar mass originates from non-supersymmetric mass terms,
the single-gravitino production rate (cf. \cite{Buchmuller:2004rq})
dominates over the pair production rate,
irrespective of the mixing with the SUSY breaking field.
}.  The relevant
couplings are~\cite{Hashimoto:1998mu,KYY,WessBagger}
\bea
   e^{-1}\mathcal{L} &=&
   - \frac{1}{8} \epsilon^{\mu\nu\rho\sigma}
   \left( G_\phi \partial_\rho \phi + G_z \partial_\rho z -  
{\rm h.c.}
     \right)
   \bar \psi_\mu \gamma_\nu \psi_\sigma\non\\
   &&
   - \frac{1}{8} e^{G/2} \left( G_\phi \phi + G_z z +{\rm h.c.}
    \right)
   \bar\psi_\mu \left[\gamma^\mu,\gamma^\nu\right] \psi_\nu,
   \label{eq:phi2gravitino}
\eea
where $\psi_\mu$ is the gravitino field, and we have chosen the
unitary gauge in the Einstein frame with the Planck units, $M_P =1$.
One has to take account of the mixing between $\phi$ and $z (z^\dag)$
discussed in the previous section, in order to evaluate the decay
rate~\cite{Dine:2006ii}.  That is to say, we should rewrite the
interactions in terms of the mass-eigenstate basis $(\tilde{\Phi},
\tilde Z)$.

To this end, we first estimate the coupling to the gravitinos,
$G_\Phi$, in the NA mass-eigenstate basis $(\Phi, \,Z)$.  In this
basis, the off-diagonal components of the non-analytic mass term
should vanish by definition:
\beq
M^2_{\Phi \bar Z} = \nabla_\Phi G_\Phi \nabla_{\bar Z} G_{\bar \Phi }
+ \nabla_\Phi G_Z \nabla_{\bar Z} G_{\bar Z} - R_{\Phi  
\bar Z i j^*}G^i G^{j^*} =0.
\eeq
Using $|\nabla_\Phi G_\Phi| \gg |\nabla_Z G_Z|$,
we obtain
\beq
\label{eq:Gzphi}
\nabla_{\bar Z} G_{\bar \Phi } \;\simeq\; \frac{R_{\Phi \bar Z i  
j^*}G^i G^{j^*} }{ \nabla_\Phi G_\Phi }.
\eeq
On the other hand, the potential minimization condition for $\Phi$ reads
\beq
G_{\bar \Phi} \nabla_\Phi G_\Phi + G_{\bar Z} \nabla_\Phi G_Z  
+ G_\Phi = 0,
\eeq
which can be solved for $G_\Phi$:
\beq
\label{eq:Gphi}
G_\Phi  \;\simeq\; -  \frac{\nabla_{\bar \Phi} G_{\bar Z}}{ \nabla_ 
{\bar \Phi} G_{\bar \Phi}} G_{Z},
\eeq
where we have used $|\nabla_\Phi G_\Phi| \gg 1$ again.  
Substituting (\ref{eq:Gzphi}) into (\ref{eq:Gphi}),
we arrive at
\beq
\label{eq:gphi-na}
|G_\Phi| \;\simeq\; 3\sqrt{3}  \frac{|R_{\Phi \bar Z Z \bar Z}| }{|  
\nabla_{\bar \Phi} G_{\bar \Phi}|^2},
\eeq
where we have used $|G_Z| = |G^Z| \simeq \sqrt{3}$.  Thus $G_\Phi$ is
always proportional to $m_{3/2}^2/m_\phi^2 \ll 1$, while it can be
enhanced if $Z$ has a quite large SUSY breaking mass, $m_z^2 \simeq 3
|R_{z \bar z z \bar z}| e^G \gg e^G$. It should be noted that
(\ref{eq:gphi-na}) always holds in the NA mass-eigenstate basis as
long as $m_\phi \gg m_{3/2}$, irrespective of the value of $m_z$.  For
the minimal K\"ahler potential, $G_\Phi$ is exactly zero in this
basis.  However one must keep in mind that $\Phi$ is generally not
identical to the true mass eigenstate $\tilde{\Phi}$.  In general, the
true mass eigenstate $\tilde{\Phi}$ ($\tilde{Z}$) is no longer a
scalar component of a chiral superfield [see Eqs.~(\ref{eq:a-phi}) and
(\ref{eq:a-z})], and hence the above consideration for $G_\Phi$ does
not hold.

Now let us write down the interactions in the mass-eigenstate basis
$(\tilde{\Phi}, \tilde{Z})$.  In the NA mass-eigenstate basis, the
couplings to the gravitinos are obtained by replacing $(\phi, z)$ with
$(\Phi, Z)$ in (\ref{eq:phi2gravitino}).  By performing the
transformation (\ref{eq:a-phi}) and (\ref{eq:a-z}), we can rewrite the
interactions in terms of $\tilde{\Phi}$ and $\tilde Z$:
\bea
   e^{-1}\mathcal{L} &\simeq&
   - \frac{1}{8} \epsilon^{\mu\nu\rho\sigma}
\left({\cal G}_\Phi^{(-)}\, \partial_\rho \tilde\Phi -
  {\cal G}_\Phi^{(+)\dag}\, \partial_\rho \tilde\Phi^\dag
+G_{Z} \partial_\rho \tilde Z - G_{\bar Z} \partial_\rho \tilde Z^\dag \right)
    \bar \psi_\mu \gamma_\nu \psi_\sigma
    \non\\&&
   - \frac{1}{8} e^{G/2} \left( {\cal G}_\Phi^{(+)}\ \tilde\Phi + 
     {\cal G}_\Phi^{(-)\dag}\ \tilde\Phi^\dag
   +G_{Z} \tilde Z + G_{\bar Z} \tilde Z^\dag \right)
   \bar\psi_\mu \left[\gamma^\mu,\gamma^\nu\right] \psi_\nu,
\eea
where we have defined
\bea
{\cal G}_\Phi^{(\pm)} &\equiv& G_\Phi \pm \tilde{\epsilon}^* G_{\bar Z}.
\eea
The decay rate of $\tilde\Phi$ is~\cite{Endo:2006zj,Nakamura:2006uc}
\begin{eqnarray}
   \Gamma(\tilde\Phi \rightarrow 2\psi_{3/2}) \;\simeq\;
   \frac{|{\cal G}_\Phi^{(eff)}|^2}{288\pi}  \frac{m_\phi^5}{m_{3/2}^2  
M_P^2},
   \label{eq:gamma_gravitino_GX}
\end{eqnarray}
for $m_\phi \gg m_{3/2}$, where we defined $|{\cal G}_\Phi^ {(eff)}|^2
\equiv 1/2\,(| {\cal G}_\Phi^{+}|^2+ | {\cal G}_\Phi^{-}|^2) =
|G_\Phi|^2 + \left| \tilde{\epsilon}^* G_{\bar Z}\right|^2$.  The
gravitino mass in the denominator arises from the longitudinal
component of the gravitino. An interpretation in the goldstino limit
is given in the Appendix.~\ref{appa}.

Let us now evaluate the order-of-magnitude of $|{\cal
  G}_\Phi^{(eff)}|^2 = |G_\Phi|^2 + \left| \tilde{\epsilon}^* G_{\bar
  Z}\right|^2$.  The first term can be related to $m_z$ if $z$ is
heavier than the gravitino due to a non-supersymmetric mass,
$m_z^2 \simeq 3 |R_{z \bar z z \bar z}| e^G  \gg m_{3/2}^2$:
\beq
|R_{\Phi \bar Z Z \bar Z}| \;\simeq\;  
\epsilon_{z \Phi} \frac{m_z^2}{3 m_{3/2}^2},
\eeq
where $\epsilon_{z \Phi}$ represents the mixing 
of $\Phi$ into $z$, and it
can be approximately given by
\beq
\epsilon_{z \Phi} \;\simeq\;\left\{
\bear{cc}
\displaystyle{ |g_{\phi \bar z}|+ |\nabla_\phi G_z| \frac{m_{3/2}}{m_\phi}}
&~~~{\rm for}~~~m_\phi \gg m_z,\\
\displaystyle{ |g_{\phi \bar z}| \frac{m_\phi^2}{m_z^2} + |\nabla_\phi G_z| \frac{m_{3/2} m_\phi }{m_z^2}}
&~~~{\rm for}~~~m_z \gg m_\phi.
\eear
\right.
\eeq
If $m_z \sim m_{3/2}$, however, $R_{\Phi \bar Z Z \bar Z}$ is not necessarily 
related to $m_z$.
On the other hand, $|\tilde \epsilon|$ is~\footnote{
Note that $\tilde \epsilon$
is  at least $O(\la \phi \ra m_{3/2}^2/m_\phi^2)$ even in the case of the minimal K\"ahler potential.
} 
\beq
|\tilde \epsilon| \;\simeq\; \left\{
\bear{cc}
\displaystyle{ \sqrt{3} \,|g_{\bar \Phi Z Z}| \frac{m_{3/2}}{m_\phi}}
&~~~{\rm for}~~~m_\phi \gg m_z,\\
\displaystyle{\sqrt{3} \,|g_{\bar \Phi Z Z}| \frac{m_{3/2}m_\phi}{m_z^2}}
&~~~{\rm for}~~~m_z \gg m_\phi.
\eear
\right.
\eeq
In summary, $|{\cal G}_\Phi^{(eff)}|^2$ is given by
\beq
\label{gphi1}
|{\cal G}_\Phi^{(eff)}|^2 \;\simeq\; 
\left| 3 \sqrt{3} \,R_{\Phi \bar Z Z \bar Z} 
\frac{m_{3/2}^2}{m_\phi^2} \right|^2 
+ \left|3\, g_{\bar\Phi Z Z}\, {m_{3/2} \over m_\phi}\right|^2
\eeq
for $m_\phi \gg m_z \sim m_{3/2}$,
\beq
\label{gphi2}
|{\cal G}_\Phi^{(eff)}|^2 \;\simeq\; 
\left|\sqrt{3}\,g_{\phi \bar z} \frac{m_z^2}{m_\phi^2} \right|^2 
+ \left|\sqrt{3} (\nabla_\phi G_z)  \frac{m_{3/2}m_z^2}{m_\phi^3}\right|^2
+ \left|3\, g_{\bar\Phi Z Z}\, {m_{3/2} \over m_\phi}\right|^2
\eeq
for $m_\phi \gg m_z  \gg m_{3/2}$, and
\beq
\label{gphi3}
|{\cal G}_\Phi^{(eff)}|^2 \;\simeq\; 
\left|\sqrt{3}\,g_{\phi \bar z} \right|^2 +
\left|\sqrt{3} (\nabla_\phi G_z) \frac{m_{3/2}}{m_\phi}\right|^2
+ \left|3 \,g_{\bar\Phi Z Z}\, {m_{3/2}m_\phi \over m_z^2}\right|^2
\eeq
for $m_z \gg m_\phi$~\footnote
{
We are grateful to M.~Ibe and Y.~Shinbara for pointing out the 1st term in Eq.(\ref{gphi3}).
}.  Note that, in the model basis,  $|\nabla_\phi G_z| \simeq O(1)$ for a modulus
field with its VEV of order  $M_P$~\cite{Endo:2006zj},
 while $|\nabla_\phi G_z| \sim \la \phi \ra$ for such scalar field $\phi$ with
 the K\"ahelr potential $K = |\phi|^2 + \cdots$ before expanding around the 
 VEV~\cite{Kawasaki:2006gs,Asaka:2006bv}.
Therefore, the second term in Eq.~(\ref{gphi3}) reproduces the partial decay 
rate of $\tilde \Phi$ into a pair of the gravitinos in 
Refs.~\cite{Endo:2006zj,Nakamura:2006uc,Kawasaki:2006gs,Asaka:2006bv}.
In addition, even in the case of $m_\phi \gg m_z$, the rate becomes
sizable if $g_{\bar \Phi Z Z}$ is order unity.

\subsubsection{SM (s)particles}

In the gravity mediation, there are non-renormalizable interactions
between the SUSY breaking field $z$ and the SM sector to induce the
soft SUSY breaking terms.  For instance, the gaugino masses are
obtained in the model frame by
\begin{eqnarray}
  \label{eq:z_gauge}
  {\mathcal L} \;=\; 
  \int d^2 \theta \, c_z \frac{z}{M_P} \, W^{(a)} W^{(a)}
  + h.c.
\end{eqnarray}
where $W^{(a)}$ is the supersymmetric field strength of the gauge
field, and $c_z$ is a coupling constant of order unity. The mixture
between the heavy scalar field $\phi$ and $z$ makes it possible for
$\tilde\Phi$ in the mass eigenstate to decay into the SM (s)particles
through the above coupling. Using (\ref{eq:rough-rel2}), the interaction 
between $\tilde \Phi$ and the SM (s)particles is given by
\begin{eqnarray}
  {\mathcal L}_{\tilde\Phi  WW}^{\rm (mix)} \;\sim\; 
  \frac{c_z}{M_P} 
  \epsilon_{z\tilde{\Phi}}
  \tilde \Phi
  \int d^2 \theta \, W^{(a)} W^{(a)}
  + h.c.
  \label{eq:gaugino-mix}
\end{eqnarray}
which leads to
\begin{eqnarray}
  \label{eq:induceddecay_gauge}
  \Gamma^{\rm (mix)}(\tilde\Phi \rightarrow gauge\ boson) \;\simeq\;
  \Gamma^{\rm (mix)}(\tilde\Phi \rightarrow gaugino) \;\simeq\;
  {3\over 2\pi}
  \lrf{N_g}{12}
\epsilon_{z\tilde{\Phi}}^2 
  |c_z|^2
  \frac{m_\phi^3}{M_P^2}
\end{eqnarray}
for $m_\phi \gg m_{3/2}$, where $N_g$ is the number of final states,
and $N_g=12$ for SU(3)$_c\times$SU(2)$_L\times$U(1)$_Y$. We notice
that the decay rate of the gaugino production is comparable to 
that of the gauge boson~\cite{Endo:2006zj,Nakamura:2006uc}.  
Note that this decay is always present as far as there is a mixing
between the $\phi$ and $z$. As we will see, it will become important
especially for the inflaton decay.

In the case of modulus decay, it also has a direct coupling to the SM
sector, such as the dilatonic coupling with the gauge supermultiplet,
${\mathcal L} = \frac{\lambda_G}{M_P} \int d^2 \theta \,\phi W^{(a)}
W^{(a)}+ {\rm h.c.}$. The decay rate through this coupling is given by
\begin{eqnarray}
  \label{eq:directdecay_gauge}
  \Gamma^{\rm (direct)}(\tilde\Phi \rightarrow gauge\ boson) \;\simeq\;
  \Gamma^{\rm (direct)}(\tilde\Phi \rightarrow gaugino) \;\simeq\;
  {3\over 2\pi}
  \lrf{N_g}{12}
  |\lambda_G|^2 \frac{m_\phi^3}{M_P^2}
\end{eqnarray}
for $m_\phi \gg m_{3/2}$. In the gravity mediation, therefore, the
direct decay of the modulus into the SM (s)particles becomes dominant
over that through mixings, as long as $|\lambda_G| \sim 1$ and $\epsilon_{z\tilde{\Phi}}
< 1$. Note that in the case of inflaton, it does not necessarily have
the above direct coupling.

\subsubsection{SUSY breaking sector}
\label{seubsec:PhiToZ}

The heavy scalar can also decay into the hidden sector, which includes
the SUSY breaking fields.  Due to the mixing between the fields $\phi$
and $z(z^\dag)$ in the model frame, the mass eigenstate $\tilde\Phi$
has a branch of the production of the hidden sector field $\tilde Z$,
if kinematically allowed. In this subsection we discuss the decay
$\tilde\Phi\to \tilde Z$ assuming $m_\phi \gg m_z$.

A possible interaction between $\phi$ and $z$ comes from the K\"ahler
potential, $K = g_{\phi \bar zz} \phi z^\dagger z + {\rm
h.c.}$. Actually such an interaction is plausible once we consider an
operator, $K = |\phi|^2|z|^2/M_P^2$, with taking account of the VEV of
$\phi$. The decay rate via this operator is however suppressed. If the
decay is induced by the $D=5$ operator in the K\"ahler potential, and
if the final states have opposite chirality, the relevant coupling
constant becomes proportional to the mass squared of the final state,
$m_Z^2$, or that of the gravitino,$m_{3/2}^2$.  Therefore the
resultant decay becomes suppressed: $\Gamma\sim
\mathrm{Max}[m_{3/2}^4,m_z^4]/m_\phi M_P^2$.

In the K\"ahler potential, there is another  $D=5$
operator, $K = g_{\bar\phi z z} \phi^\dagger zz + {\rm h.c.}$, which
is different from the above one: the final states have the same
chirality. This operator induces a larger decay rate:
\begin{eqnarray}
  \label{eq:decay_SUSYbreaking}
  \Gamma(\tilde\Phi \rightarrow \tilde Z\tilde Z) \;\simeq\;
  \frac{|g_{\bar\phi z z}|^2}{8\pi} \frac{m_{\phi}^3}{M_P^2}.
\end{eqnarray}
As discussed in the previous subsection, the K\"ahler mixing
$g_{\bar\phi z z}$ also induces the decay into the gravitino
[cf. (\ref{eq:gamma_gravitino_GX}), (\ref{gphi1}), and (\ref{gphi2})].
It should be noted that these two decay rates can be correlated. As
long as $|{\cal G}_\Phi^{(eff)}|$ is dominated by the last term in
(\ref{gphi1}) or (\ref{gphi2}) that contains $|g_{\bar \Phi Z Z}|$,
they become comparable: $\Gamma(\tilde{\Phi}\to
2\psi_{3/2})/\Gamma(\tilde{\Phi}\to 2\tilde{Z})\simeq 1/4$.

It is stressed that production of the fermionic component of the $z$
field is very different. This is because a combination of the
fermionic components of $\phi$ and the SUSY breaking fields is
absorbed into the gravitino as a goldstino. In the minimum setup,
namely where there is a single SUSY breaking field, the fermionic
component of $z$ almost behaves as the goldstino, and that of $\phi$
provides the remnant massive degrees of freedom. Therefore, when the
$\phi$ mass is given by the supersymmetric term, $\nabla_\phi G_\phi$,
the mass of this massive fermion becomes close to $m_\phi$, and hence
the decay is kinematically suppressed or forbidden.

The produced $\tilde{Z}$ subsequently decays into the visible sector,
and into the gravitino if kinematically allowed. The decay of the
$\tilde{Z}$ field and its implications will be discussed in the
following sections.

\subsection{Modulus}

In this and the next sections, we discuss how the decay via mixings
with the SUSY breaking field affect the cosmological scenarios.  We
concentrate especially on the modulus and the inflaton, and see how
disastrous the cosmological scenarios become due to such mixings.

Let us start from the modulus decay. We discuss two distinct cases
$m_z > m_\phi$ and $m_\phi < m_z$ in turn.  In both cases, the
dominant decay channel is that into the SM (s)particles, whose rate is
given by Eq.~(\ref{eq:directdecay_gauge}). A successful big bang nucleosynthesis (BBN) requires a
temperature higher than $\sim
5$~MeV~\cite{Kawasaki:1999na,Hannestad:2004px}, which leads to a lower
bound on the modulus mass, $m_\phi \gsim 100$~TeV, and we assume this
is the case.

Here, we should mention that there may be another cosmological moduli
problem associated with the SUSY breaking field $z$.  In this section
we assume that it is the heavy $\phi$ field which dominates the energy
density of the universe and causes the final reheating (before the
BBN), and we will briefly discuss the $z$ modulus problem in
Sec.~\ref{sec:5}.

\subsubsection{$m_z > m_\phi$}

In this case, the modulus field $\phi$ decays into the SM (s)particles
and the gravitino, with the partial decay rates in
Eq.~(\ref{eq:directdecay_gauge}) and
Eq.~(\ref{eq:gamma_gravitino_GX}), respectively.  The
branching ratio of the gravitino production then becomes
\begin{eqnarray}
  B_{3/2} = Br(\tilde{\Phi}\to 2\psi_{3/2}) &=&
  {1\over 72 N_g |\lambda_G|^2}
  {m_\phi^2 \over m_{3/2}^2}
  |{\cal G}_\Phi^{(eff)}|^2
\end{eqnarray}
As can be seen in Eqs.~(\ref{gphi1}) - (\ref{gphi3}), 
the coupling $|{\cal G}_\Phi^{(eff)}|$
depends on the model.  If $m_z > m_\phi$, however, 
$|{\cal G}_\Phi^{(eff)}|$ is suppressed 
only by a single power of the gravitino mass 
and the branching ratio becomes
\begin{eqnarray}
  B_{3/2} &\simeq&
  {1\over 24 N_g |\lambda_G|^2}
 \left( |\nabla_\phi G_z|^2 + |g_{\phi \bar z}|^2 \frac{m_\phi^2}{m_{3/2}^2}\right).
\end{eqnarray}
Using $ |\nabla_\phi G_z| \sim O(1)$ in the model basis~\cite{Endo:2006zj}, 
the first term is the same order as the one estimated in
Refs.~\cite{Endo:2006zj,Nakamura:2006uc}. As was shown there, such a
large branching fraction of the gravitino production causes serious
cosmological problems. The second term makes the problem even worse
if $|g_{\phi \bar z}| \gg m_{3/2}/m_\phi$.

\subsubsection{$m_\phi > m_z$}

Now one has to consider a new decay mode, $\tilde{\Phi} \to
2\tilde{Z}$, in addition to the channels discussed above. As discussed
in Sec.\ref{seubsec:PhiToZ}, if $g_{\bar \phi z z}$ is sizable, 
the $\tilde{\Phi}$ produces roughly as many
$\tilde{Z}$ as the gravitino. Here we discuss the fate of the produced
$\tilde{Z}$ and its implications.

If $m_z \lsim 2\,m_{3/2}$, the $\tilde{Z}$ dominantly decays into the
visible sector via the interaction Eq.~(\ref{eq:z_gauge}), which leads
to $\Gamma(\tilde{Z}\to visible)\sim m_z^3/M_P^2$.  Note that this
rate is comparable to the decay rate of the gravitino for $m_z\sim
m_{3/2}$. Therefore it causes qualitatively similar problems as the
gravitino, such as changing light element abundances and producing too
many LSPs~\cite{Endo:2006zj,Nakamura:2006uc}.  The details of the
constraint on the model depends on the mass and couplings of the $z$
field.

If on the other hand $m_z \gg m_{3/2}$, due to SUSY breaking mass
term, $\tilde{Z}$ dominantly decays into the gravitino. (Note that
$|G_Z|\simeq \sqrt{3}$ and the rate is enhanced by
$(m_z/m_{3/2})^2$. cf. Eq.~(\ref{eq:gamma_gravitino_GX}).)  Recall that
there are gravitinos directly produced by the $\Phi$ decay. The net
effect is therefore just an enhancement of the gravitino abundance by
an order one factor.  The subsequent gravitino decay is subject to the
cosmological constraints~\cite{Endo:2006zj,Nakamura:2006uc}.

To summarize, $\tilde{\Phi}$ produces roughly as many
$\tilde{Z}$ as the gravitinos, and the produced $\tilde{Z}$ will cause
a similar problem as the gravitino does.

\subsection{Inflaton}

We now turn to discuss the inflaton decay.
We assume that the SUSY breaking field $z$ is lighter than the
inflaton $\phi$~\footnote{Note, however, that this may not be the case in 
the new inflation models~\cite{Kawasaki:2006gs}.}. 
Therefore, the inflaton can decay into the SM (s)particles,
gravitinos, and the SUSY breaking sector fields. The importance of the
inflaton decay into the gravitino has been recently pointed
out in Ref.~\cite{Kawasaki:2006gs}.

Let us first consider the inflaton decay into the SM (s)particles through the
interaction (\ref{eq:gaugino-mix}). The mixing with the SUSY breaking field may
enhance the decay rate of the inflaton, which leads to a higher reheating 
temperature, $T_R$. Since $T_R$ is bounded from above due to the abundance 
of the gravitinos produced by thermal scatterings,  such mixing must be 
small enough. 

The presence of the interaction  (\ref{eq:gaugino-mix}) sets
 a lower bound on the reheating temperature:
\beq
 T_R \gtrsim 3 \times 10^8~{\rm GeV}\, 
 \epsilon_{z \tilde \Phi} |c_z| \lrfp{m_\phi}{10^{12}{\rm GeV}}{\frac{3}{2}}
\label{eq:tr-mix}			  
\eeq
where we have used $N_g = 12$
for the SM gauge groups and  the relativistic degrees of freedom 
$g_* \simeq 200$. For $m_{3/2} \simeq  O(0.1 -1{\rm \,TeV})$,  
the bound from the gravitino problem
reads $T_R <  O(10^{6} - 10^{8})\,$GeV~\cite{Kawasaki:2004yh,Kohri:2005wn}, 
where the upper bound depends on the gravitino mass and the hadronic 
branching ratio $B_h$. Combining this with (\ref{eq:tr-mix}), we obtain
\beq
\label{eq:ep-gra}
\epsilon_{z \tilde \Phi} \lesssim (3\times10^{-3}- 0.3) \,c_z^{-1} 
\lrfp{m_\phi}{10^{12}{\rm GeV}}{-\frac{3}{2}}.
\eeq
The heavier the inflaton mass is, the severer this bound becomes.  For
the new inflation model~\cite{Kumekawa:1994gx,Izawa:1996dv}, the
inflaton mass is relatively small, $m_\phi \sim 10^{10}$~GeV, and
therefore the bound (\ref{eq:ep-gra}) does not give any sensible
constraint on the mixing. For the hybrid inflation
models~\cite{Copeland:1994vg,Dvali:1994ms,Linde:1997sj}, however, the
inflaton mass can be very large, $m_\phi \sim O(10^{11} -
10^{15})$~GeV~\footnote {
The hybrid inflation models contain two types of the fields: the
inflaton field and the waterfall fields. Although the bound on
$|g_{\phi \bar z}|$ applies to both fields, we identify $\phi$ with
the waterfall field when we substitute the VEV of $\phi$ into
(\ref{eq:g-phi-barz}).
}
Then we obtain $\epsilon_{z \tilde \Phi}  \lesssim O(10^{-7} - 10)$. 
To translate this bound
into that on parameters in the model basis, let us estimate 
$\epsilon_{z \tilde \Phi}$,
\beq
\label{eq:ep-z-phi}
\epsilon_{z \tilde \Phi} \simeq {\rm Max} 
\left[|g_{\phi \bar z}|, \sqrt{3} \phi_0 \frac{m_{3/2}}{m_\phi}, 
\sqrt{3} |g_{\bar \phi z z}| \frac{m_{3/2}}{m_\phi} \right],
\eeq
where we assumed $m_\phi \gg m_z$. Since the second and third 
terms are highly suppressed
due to the ratio of the gravitino mass to the inflaton mass, 
 the bound on $\epsilon_{z \tilde \Phi}$ is effectively 
  that on $|g_{\phi \bar z}|$. In the case of the hybrid inflation 
model, therefore,
we obtain a nontrivial bound, $|g_{\phi \bar z}| \lesssim O(10^{-7} - 10)$.

To see how severe the bound on the mixing is, it is necessary to 
consider explicit interactions
in the K\"ahler potential~\footnote{
We assume here that $\phi$ and $z$ are not coupled 
in the superpotential for simplicity.
}. Let us consider the following interactions in the model 
basis before expanding the fields 
around their VEVs,
\beq
\delta K = k_1 |\phi|^2 (z+z^\dag) + k_2 |\phi|^2 |z|^2 
+\frac{ k_3}{2} |\phi|^2 (zz+z^\dag z^\dag) \cdots,
\eeq
where $k_i~(i=1,2,3)$ are numerical coefficients, and we 
have dropped several terms like
$\phi^2 (z + z^\dag)$, assuming that $\phi$ is charged under some symmetry. 
As long as $z$ is a singlet, all the coefficients are expected 
to be order unity. 
Then $g_{\phi \bar z}$ is non-vanishing if $\phi$ and $z$ take non-zero VEVs,
\beq
\label{eq:g-phi-barz}
g_{\phi \bar z} = k_1 \phi_0^* + k_2 \phi_0^* z_0 + k_3 \phi_0^* z_0^*.
\eeq
Therefore the constraint on $g_{\phi \bar z}$ can be interpreted as that on the
numerical coefficients  $k_i$, which are otherwise unconstrained 
from any symmetries of $\phi$.
If $k_i$ is severely constrained  from cosmological considerations, 
it indicates either
that there is still unknown symmetry or mechanism to suppress the 
couplings, or that such inflation model with vanishing $\phi_0$ is favored. 
As an example, let us take the hybrid inflation model with $\phi_0 \sim 10^{-3}$. 
Then $|g_{\phi \bar z}| \lesssim O(10^{-7} - 10)$
can be rephrased as $|k_1 + k_2   z_0 + k_3 z_0^*| \lesssim O(10^{-4} - 10^4)$.
Therefore,  a considerable part of the parameter spaces are disfavored 
if $B_h \simeq 1$.
(Note, however, that the hybrid inflation model is already disfavored only from 
the direct gravitino production~\cite{Kawasaki:2006gs}.)

Let us now consider the inflaton decay into the gravitinos.  Recently,
it was pointed out in Ref.~\cite{Kawasaki:2006gs} that the gravitino
production from the inflaton decay can put a severe constraint on the
inflation models (in particular, the hybrid inflation model is
excluded unless the higher order terms in the K\"ahler potential is
extremely suppressed). The decay rate into a pair of the gravitinos is
given by (\ref{eq:gamma_gravitino_GX}).  The constraint on the
inflation models can be read from Fig.~1 in
Ref.~\cite{Kawasaki:2006gs} by replacing $G_\phi$ with $|{\cal
G}_\Phi^{(eff)}| \simeq |3 g_{\bar \Phi Z Z} \,m_{3/2}/m_\phi|$.
Thus, we can rephrase the results of Ref.~\cite{Kawasaki:2006gs} that
the hybrid inflation model is excluded unless $|g_{\bar \Phi Z Z}|$ is
extremely small in the gravity mediation.

Lastly, let us consider the inflaton decay into the SUSY breaking
sector.  As in the case of the moduli, the inflaton decay into $\tilde
Z$ is always concomitant with almost same amount of the gravitino
production, since the both production rates are proportional to
$|g_{\bar \Phi Z Z}|^2$. Therefore the produced $\tilde Z$ only causes
a problem which is at most as severe as the gravitino overproduction
problem.

\section{Low Energy SUSY Breaking Models}
\label{sec:4}

In this section we consider low energy SUSY breaking models, as
represented by the GMSB models. Compared to the gravity mediation,
there are two major differences. One is that the SUSY breaking field
couples to the visible sector more strongly, which is a general
feature of the low energy SUSY breaking models.  This enhances the
decay rate of $\tilde \Phi$ due to the mixing. The other is the
existence of the messenger sector fields, which is characteristic to
the GMSB models.  Since the messenger sector contains another scalar
field, $s$, we need to consider the scalar mixings of both $\phi - z$
and $\phi - s$.

In the messenger sector, there is a chiral superfield, $s$, with  nonzero
VEVs of the scalar and auxiliary components, which couples to the messengers 
$\Psi_M$ and  $\bar \Psi_M$  by
\beq
\label{eq:gmsb}
W \;=\; y_M\,s \,\Psi_M \bar \Psi_M,
\eeq
where $y_M$ is a coupling constant. The scalar VEV, $M_s \equiv \la s
\ra$, sets the messenger mass scale, $M_{\rm mess} \equiv y_M M_s$,
while the F-term, $F_s$, provides the mass splitting between the
messenger fermions and bosons, $\sim y_M F_s$.  The SUSY breaking is
transmitted radiatively to the visible sector by the SM gauge
interactions, under which $\Psi_M$ and $\bar \Psi_M$ are charged.  The
SUSY breaking scale in the visible sector is therefore determined by
$F_{s} / M_s$.  For example, the gaugino mass is induced by~\footnote{
Note that such an interaction as (\ref{eq:s_gauge}) always exists 
in the low energy
SUSY breaking models, even if the messenger sector does not exist.
In this case $M_s$ simply parametrizes the strength of the interaction between
$s$ and the visible sector. 
}
\begin{eqnarray}
  \label{eq:s_gauge}
  {\mathcal L} \;=\; 
  \int d^2 \theta \, \frac{\alpha}{4\pi} \frac{s}{M_s} \, W^{(a)} W^{(a)}
  + h.c.,
\end{eqnarray}
where $\alpha$ denotes the gauge coupling, and we have assumed the messenger index $N=1$.
The gaugino mass $M_\lambda$ is 
therefore given by 
\beq
M_\lambda = {\alpha \over 4\pi} \,{F_{s} \over M_s}.
\eeq
By using $G_s = F_s/(m_{3/2}M_P)$, we can relate $M_s$ to $m_{3/2}$:
\beq
\label{mg-ms}
m_{3/2} \simeq 9 \times 10^{-4} {\rm\,GeV}\,G_s^{-1} 
\lrf{m_{\tilde g}}{1{\rm\,TeV}}
\lrf{M_s}{10^{10}{\rm \,GeV}} ,
\eeq
where the gluino mass $m_{\tilde g}$ is determined at the messenger
scale.  In contrast to the gravity mediation, it is nontrivial (and
therefore model-dependent) how large $G_s$ is. In fact, in the direct
gauge-mediation scenario, $|G_s| \sim 1$ if $s$ is identified with
$z$, while $|G_s| \ll 1$ in such models that the SUSY breaking effects
is radiatively transmitted from a secluded sector (that contains $z$)
to the messenger sector. If $|G_s| \sim 1$, there is no significant
difference between $s$ and $z$.  If $|G_s| \ll 1$, we need to consider
the mixings $\phi - z$ and $\phi - s$, separately (for simplicity we
neglect the mixing between $z$ and $s$). The formalism developed in
Sec.~\ref{sec:2} can also be applied to the $\phi - s$ mixing.  Since
$|G_z| \sim \sqrt{3}$, it is the mixing with $z$ that determines the
decay of $\phi$ into the gravitinos. On the other hand, it is $s$ that
determines the decay into the SM (s)particles, since $s$ (not $z$)
couples to the SM (s)particles via the messengers $\Psi_M$ and $\bar
\Psi_M$.  Lastly $\phi$ may decay into both $s$ and $z$ via the
mixings. Thus, although there are two SUSY breaking fields $s$ and $z$
in the GMSB models, we can similarly discuss the decay of $\phi$ as we
did in Sec.~\ref{sec:3}.

\subsection{Decay Modes}
The decay channels of the heavy scalar $\tilde\Phi$ are quite similar to
those in the gravity mediation. In particular, the gravitino
production rate is independent of the couplings between the SUSY
breaking field and the visible sector. In addition, the decays of
$\tilde\Phi$ into $z$ and $s$ (if kinematically allowed) are also
similar to the gravity mediation case, i.e., they are dominated by the
decays induced by the higher order couplings in the K\"ahler potential
$g_{\bar \phi z z}$ and $g_{\bar \phi s s}$, respectively, if these
couplings are sizable, and otherwise suppressed. Here, we focus on the
new features of the low energy SUSY breaking scenario.

When the SUSY breaking effects are mediated to the visible sector 
with the interactions with a lower cutoff, $M_s$, the mixing-induced 
$\tilde\Phi$ decay into the visible sector depends on $M_s$ rather 
than $M_P$. Since the field $s$ couples with the visible sector, the 
decay rate is evaluated from the operator Eq.~(\ref{eq:s_gauge}) as
\begin{eqnarray}
\label{eq:decay-rate-th-mix}
  \Gamma^{\rm (mix)}(\tilde\Phi \rightarrow gauge\ boson) \;\simeq\;
  \Gamma^{\rm (mix)}(\tilde\Phi \rightarrow gaugino) \;\simeq\;
   {\alpha_s^2 \over 16\pi^3}
  \lrf{N_g}{8}
  \epsilon_{s\tilde{\Phi}}^2 
  \frac{m_\phi^3}{M_s^2},
\end{eqnarray}
where $\epsilon_{s\tilde{\Phi}}$ is defined as
Eq.~(\ref{eq:rough-rel2}) with $z \rightarrow s$, and we assumed the
decay is dominated by the gluon/gluino production.  The decay rate
thus depends on the mixing, $\epsilon_{s\tilde{\Phi}}$. If the mixing
is dominated by the non-renormalizable term in the K\"ahler potential,
$K = |\phi|^2 |s|^2/M_P^2$, $\epsilon_{s\tilde{\Phi}}$ is given by
$\sim M_s/M_P$.  Then the mixing-induced decay rate becomes $\Gamma
\sim m_\phi^3/M_P^2$.

In the GMSB setups, there exist the messenger fields, $\Psi_M$ and
$\bar \Psi_M$. Since they interacts with the $s$ field by the
renormalizable coupling, Eq.~(\ref{eq:gmsb}), the heavy scalar field,
$\tilde\Phi$, can decay into the fermionic component of $\Psi_M$ and
$\bar \Psi_M$ rapidly as long as the channel is kinematically
allowed. From Eq.~(\ref{eq:gmsb}), the decay rate is estimated as
\begin{eqnarray}
\label{eq:gamma-mes}
  \Gamma(\tilde\Phi \rightarrow \psi_\Psi \psi_{\bar\Psi})
  \;\simeq\; N_{\rm mess} \frac{|y_M \epsilon_{s \tilde \Phi}|^2}{32\pi} 
  m_\phi,
\end{eqnarray}
where $N_{\rm mess}$ is a number of the possible final states, for
instance, $N_{\rm mess} = 5$ when $\Psi_M + \bar\Psi_M$ are charged as
$\mathbf{5} + \mathbf{\bar 5}$ under $SU(5)_{\rm GUT}$. Therefore
unless $y_M$ and/or $\epsilon_{s\tilde{\Phi}}$ is extremely
suppressed, the dominant channel of $\tilde\Phi$ becomes the
production of the messenger fermion.

\subsection{Modulus}
The modulus decay in the low energy SUSY breaking scenario is similar
to that in the gravity mediation, as long as $\epsilon_{s\tilde{\Phi}}
\lesssim M_s/M_P$. This is the case if the mixing mainly comes from
e.g., $\delta K = \kappa\, |\phi|^2 |s|^2/M_P^2$. The modulus decay
into the SM (s)particles via the mixing with $s$ then proceeds with
the rate (\ref{eq:decay-rate-th-mix}) that is at most comparable to
(\ref{eq:directdecay_gauge}). Therefore a successful BBN requires
$m_\phi \gtrsim 100$ TeV as in the case of gravity mediation.  In the
general low energy SUSY breaking models, however, $s$ can be a singlet
field and therefore such an interaction as $\delta K = |\phi|^2
(s+s^\dag)/M_P$ may exist~\footnote{Note that this coupling enhances the gravitino production rate due to the first term in Eq.~(\ref{gphi3}) if $G_s$ is
sizable.  }. 
In this case, the modulus decay into the
visible sector via the mixing can exceed the rate
(\ref{eq:directdecay_gauge}), and the modulus may decay before the BBN
even if its mass $m_\phi$ is smaller than $100$~TeV. Although this may
relax the moduli problem in the low energy SUSY breaking models, it
strongly depends on the nature of $s$ whether such an interaction
exists at all.

Actually, there may be a cosmological moduli problem associated with
the $z$ and/or $s$ fields. Here, we discuss the case where the
universe is dominated and reheated by the heavy field $\phi$, and
leave the other cases for discussion in Sec.~\ref{sec:5}.

The gravitino production occurs via the mixing of $\phi$ with $z$ (and
$s$ if $G_s \sim O(1)$ as in the direct gauge mediation), and the
decay rate is given by (\ref{eq:gamma_gravitino_GX}). The cosmological
constraints on the stable gravitinos from the modulus decay are as
given in Ref.~\cite{Endo:2006zj}.

If kinematically allowed, and if $g_{\bar \phi z z}$ and $g_{\bar \phi
s s}$ are non-vanishing, the modulus decays into $z$ and $s$. Although
the masses of $z$ and $s$ are considered to be comparable or larger
than $m_{3/2}$, they are model-dependent. So here we take $m_z$ and
$m_s$ as free parameters.  The abundance of $z$ is the same order of
the gravitino abundance if $g_{\bar \phi z z}$ is sizable, just as the
case in the gravity mediation.  On the other hand, unless $G_s \sim
O(1)$, the $s$ abundance is not necessarily correlated with the
gravitinos.  For $m_z > 2 m_{3/2}$, the produced $z$ can decay into a
pair of gravitinos, and the rate is enhanced for larger $m_z$.  The
$z$ field may decay into the SM (s)particles, if it has a direct
coupling to the visible sector. The strength of the coupling must be
such that the $F$-term of $z$ does not give dominant contributions to
the soft masses. In addition, it is possible that $z$ has relatively
strong interactions with $s$ and decays into $s$.
Then we only have to consider the decay
processes of $s$, which comes both directly from the modulus decay and
through the decay of $z$. The interaction of $s$ with the visible
sector is given by (\ref{eq:s_gauge}).  Assuming that $s$ dominantly
decays into two gluons, the decay temperature is given by
\beq
T_d^{(s)} \;\simeq\; 0.04 {\rm\,GeV} 
\lrfp{m_s}{10{\rm\,GeV}}{\frac{3}{2}} \lrfp{M_s}{10^{10}{\rm\,GeV}}{-1},
\eeq
where we take $g_*=10.75$.
To be conservative,  we require that $s$ decays before the BBN starts, 
i.e., $T_d^{(s)} \gtrsim 5$~MeV~\cite{Kawasaki:1999na,Hannestad:2004px}.
Then the mass of $s$ must satisfy
\beq
\label{eq:ineq}
m_s \;\gtrsim\;3{\rm\,GeV} \lrfp{M_s}{10^{10}{\rm\,GeV}}{\frac{2}{3}}.
\eeq
It should be noted however that, even if this inequality is satisfied,  
$s$ may produce the too many LSPs and/or gravitinos, if kinematically allowed.

Lastly let us comment on the modulus decay into the messengers.
Although there exist the messenger fields in the GMSB models, it is
unlikely that the modulus decays into them, since the messenger scale
$M_{\rm mess}$ is typically larger than the modulus mass.

\subsection{Inflaton}

The low energy SUSY breaking models may contain the messenger sector
as in the GMSB models.  If the messenger scale $M_{\rm mess} = y_M
M_{s}$ is smaller than the inflaton mass $m_\phi$, the inflaton can
decay into the messenger sector as well via the $\phi - s$ mixing. In
the following we discuss the cases with and without such a channel
separately.

\subsubsection{Decay into visible sector, the gravitinos, $s$ and $z$}

Let us first consider the case without the decay into the messenger
sector.  The inflaton then decays into the SM (s)particles, the
gravitinos, $s$ and $z$. In the following we assume $m_\phi \gg
m_s,\,m_z$.

The decay into the SM (s)particles may proceed via the mixing with
$s$.  The interaction (\ref{eq:s_gauge}) sets a lower bound on the
reheating temperature:
\beq
\label{tr-lesb}
T_R \gtrsim 2 \times 10^{11}{\rm\,GeV} \,
\frac{\epsilon_{s \tilde\Phi}}{|G_s|} \lrf{m_{\tilde g}}{1{\rm\,TeV}}
\lrfp{m_{3/2}}{1{\rm\,GeV}}{-1} \lrfp{m_\phi}{10^{12}{\rm\,GeV}}{\frac{3}{2}},
\eeq
where we set $N_g = 8$ and $g_* \simeq 200$.  In the low energy SUSY
breaking models, the upper bound on $T_R$ is given 
by~\cite{Moroi:1993mb,Bolz:2000fu}
\beq
\label{eq:tr1}
    T_R ~\lesssim ~5\times 10^7~{\rm GeV}
 \lrfp{m_{\tilde g}}{1{\rm\,TeV}}{-2} \left(\frac{m_{3/2}}{1\,{\rm GeV}}\right),
\eeq
for $m_{3/2} =10^{-4} - 10~{\rm GeV}$, and
\beq
\label{eq:tr2}
T_R \lesssim O(100){\rm GeV},
\eeq
for $1$~keV $ \lesssim m_{3/2} \lesssim 10^{-4}$~GeV, in order for the
gravitino abundance not to exceed the dark matter abundance.  Here and
in what follows we neglect the difference of the values of $m_{\tilde
g}$ at the messenger scale and at the reheating temperature. In the
GMSB model, the assumption $M_{\rm mess} = y_M M_s > m_\phi$ sets a
lower limit on the gravitino mass.  Although the inflaton mass
$m_\phi$ strongly depends on the inflation models, it is typically
larger than $O(10^9)$~GeV.  Using (\ref{mg-ms}), the gravitino mass
should be larger than $O(10^{-4})$~GeV in this case. For the low
energy SUSY breaking models without the messenger sector, such a lower
limit is not necessarily applied.  Combining (\ref{tr-lesb}) with
(\ref{eq:tr1}) or (\ref{eq:tr2}), we obtain the severe bound on the
mixing:
\beq
\label{ep-s-phi-1}
\epsilon_{s \tilde\Phi} \lesssim 3 \times 10^{-10} |G_s| 
\lrfp{m_{\tilde g}}{1{\rm\,TeV}}{-3}
\lrfp{m_{3/2}}{1{\rm\,MeV}}{2} \lrfp{m_\phi}{10^{12}{\rm\,GeV}}{-\frac{3}{2}},
\eeq
for $m_{3/2} =10^{-4} - 10~{\rm GeV}$ and
\beq
\label{ep-s-phi-2}
\epsilon_{s \tilde\Phi} \lesssim
O(10^{-15})
 |G_s|  \lrfp{m_{\tilde g}}{1{\rm\,TeV}}{-1}
\lrf{m_{3/2}}{10^{-5}{\rm\,GeV}} 
\lrfp{m_\phi}{10^{12}{\rm\,GeV}}{-\frac{3}{2}},
\eeq
for $1$~keV $ \lesssim m_{3/2} \lesssim 10^{-4}$~GeV. The bounds
becomes severer for smaller $G_s$, larger $m_\phi$, and smaller
$m_{3/2}$.

To exemplify how severe the bound is, let us rewrite this bound to that on 
coefficients of the higher order interactions in the K\"ahler potential.  We
consider the following interactions in the model basis before expanding
around the VEVs:
\beq
\label{eq:phi-s-4}
\delta K \;=\;  k_1|\phi|^2(s+s^\dag) + k_2 |\phi|^2 |s|^2,
\eeq
where $k_1$ and $k_2$ are numerical coefficients. 
The first term can be forbidden if $s$ has some symmetries (i.e., $k_1=0$),
but we include it here to see how severely such an interaction is constrained. 
On the other hand, $k_2$ is unconstrained from any symmetries, so it is expected 
to be order unity. This K\"ahler potential leads to
\beq
g_{\phi \bar s} \;=\; k_1\, \phi_0^* + k_2\, \phi_0^* M_s,
\eeq
where we have taken the VEV of $s$ real, for simplicity.
Barring cancellations,
 we obtain the constraints on $k_1$ and $k_2$ from (\ref{ep-s-phi-1}) and
(\ref{ep-s-phi-2}), since $\epsilon_{s \tilde\Phi}$ is roughly equal to $|g_{\phi \bar s}|$ for
$m_\phi \gg m_s$:
\bea
|k_1|  &\lesssim& 8 \times 10^{-7}\, |G_s| \lrfp{m_{\tilde g}}{1{\rm\,TeV}}{-3}
\lrfp{m_{3/2}}{1{\rm\,MeV}}{2} \lrfp{m_\phi}{10^{12}{\rm\,GeV}}{-\frac{3}{2}} 
\lrfp{|\phi_0|}{10^{15}{\rm\,GeV}}{-1},\non\\
|k_2| &\lesssim& 2  \times 10^{2} \lrfp{m_{\tilde g}}{1{\rm\,TeV}}{-2}
\lrf{m_{3/2}}{1{\rm\,MeV}} \lrfp{m_\phi}{10^{12}{\rm\,GeV}}{-\frac{3}{2}} 
\lrfp{|\phi_0|}{10^{15}{\rm\,GeV}}{-1},
\label{g-s-phi-1}
\eea
for $m_{3/2} =10^{-4} - 10~{\rm GeV}$ and
\bea
|k_1| &\lesssim& O(10^{-9})  |G_s|
\lrfp{m_{\tilde g}}{1{\rm\,TeV}}{-1}
\lrf{m_{3/2}}{1{\rm\,MeV}}
\lrfp{m_\phi}{10^{12}{\rm\,GeV}}{-\frac{3}{2}} 
\lrfp{|\phi_0|}{10^{15}{\rm\,GeV}}{-1}, \non\\
|k_2| &\lesssim& O(0.1)  
\lrfp{m_\phi}{10^{12}{\rm\,GeV}}{-\frac{3}{2}} 
\lrfp{|\phi_0|}{10^{15}{\rm\,GeV}}{-1},
\label{g-s-phi-2}
\eea
for $1$~keV $ \lesssim m_{3/2} \lesssim 10^{-4}$~GeV, where we have
used (\ref{mg-ms}).  The bounds become severer for larger inflaton VEV
and mass. It should be noted that  the coefficient $k_1$ is tightly constrained,
which strongly disfavors the existence of a singlet field $s$ 
in the low energy SUSY breaking model. In other words, the $s$ field
must be charged under some symmetry (e.g. $U(1)$ symmetry) to forbid such
interaction as $\sim |\phi|^2 (s+s^\dag)$. Let us now focus on the
constraint on $k_2$, which is generically unsuppressed.
As an example, let us consider the hybrid inflation model.
For $m_{3/2} >10^{-4} {\rm \, GeV}$, only some fraction of the
parameter space is disfavored, while fairly wider ranges of the model
parameters require fine-tunings on the higher order interactions in
the K\"ahler potential for $m_{3/2} < 10^{-4}{\rm\, GeV}$.

Next let us consider the $\tilde \Phi$ decay into the gravitinos.
Since the decay is induced by the mixing with $z$ with $G_z \sim 1$,
the gravitino overproduction problem is similar to that considered in
the previous section.  The only difference is the upper bound on $T_R$
from the abundance of the gravitinos produced by thermal scatterings
[cf. (\ref{eq:tr1}) and (\ref{eq:tr2})].  Although the upper bound on $T_R$
depends on $m_{3/2}$ for $m_{3/2} >10^{-4} {\rm \, GeV}$, that on the
abundance of the gravitinos directly produced by the inflaton decay does not
depend on $m_{3/2}$. Therefore the gravitino overproduction problem
sets a more or less similar bounds on the inflation models given in
Ref.~\cite{Kawasaki:2006gs}.

Lastly let us consider the decay of $\tilde \Phi$ into the SUSY
breaking sector, $s$ and $z$.  As discussed in Sec.~\ref{sec:3}, the
decay rate into $z$ is comparable to the gravitino
production. However, in contrast to the gravity mediation, $z$ may
have relatively strong coupling with the messenger sector or the
visible sector. If this coupling is so strong that $z$ decays mainly
into the visible sector before the BBN, $z$ may not be cosmologically
problematic.  Even if the coupling is weak, $z$ decays into the
gravitinos as far as $m_z > 2 m_{3/2}$, and it only increases the
gravitino abundance by $O(1)$ factor.  Although $s$ is also produced
from the $\tilde \Phi$ decay if $g_{\bar \phi s s}$ is sizable, it
does not cause any cosmological difficulties if the inequality
(\ref{eq:ineq}) is satisfied.

\subsubsection{Decay into the messenger sector}

If there is a messenger sector as in the GMSB scenario, and if the
messenger scale $M_{\rm mess} = y_M M_{s}$ is smaller than the
inflaton mass, the inflaton can decay directly into the messenger
sector. Indeed, such a decay may make the reheating temperature of the
inflation even higher than that discussed in the previous subsection.
Using the decay rate (\ref{eq:gamma-mes}), the reheating temperature
becomes
\beq
T_R \gtrsim 2 \times 10^{14}{\rm\, GeV} \, |y_M \,\epsilon_{s \tilde \Phi}| 
\lrfp{m_\phi}{10^{12} {\rm\,GeV}}{\frac{1}{2}},
\eeq
where we set $N_{\rm mess} = 5$. Combined with (\ref{eq:tr1}) or (\ref{eq:tr2}), 
we obtain
\beq
\epsilon_{s \tilde \Phi} \;\lesssim\; 3 \times 10^{-10} |y_M|^{-1} 
\lrfp{m_{\tilde g}}{1{\rm\,TeV}}{-2}
\lrf{m_{3/2}}{1{\rm\,MeV}} 
\lrfp{m_\phi}{10^{12} {\rm\,GeV}}{-\frac{1}{2}}
\eeq
for $m_{3/2} =10^{-4} - 10~{\rm GeV}$ and
\beq
\epsilon_{s \tilde \Phi} \;\lesssim\; O(10^{-13})   |y_M|^{-1}
\lrfp{m_\phi}{10^{12} {\rm\,GeV}}{-\frac{1}{2}}
\eeq
for $1$~keV $ \lesssim m_{3/2} \lesssim 10^{-4}$~GeV.

Assuming that the mixing is provided by the interaction (\ref{eq:phi-s-4}),
we can rewrite the above bounds as
\bea
|k_1|& \lesssim& 8 \times 10^{-7} \,|y_M|^{-1}
\lrfp{m_{\tilde g}}{1{\rm\,TeV}}{-2}
\lrf{m_{3/2}}{1{\rm\,MeV}}  \lrfp{m_\phi}{10^{12}{\rm\,GeV}}{-\frac{1}{2}} 
\lrfp{|\phi_0|}{10^{15}{\rm\,GeV}}{-1},\non\\
|k_2|& \lesssim& 2 \times 10^{2} \,
\lrfp{m_{\tilde g}}{1{\rm\,TeV}}{-2}
\lrf{m_{3/2}}{1{\rm\,MeV}}  \lrfp{M_{\rm mess}}{10^{10}{\rm\,GeV}}{-1}
\lrfp{m_\phi}{10^{12}{\rm\,GeV}}{-\frac{1}{2}} 
\lrfp{|\phi_0|}{10^{15}{\rm\,GeV}}{-1},
\label{k-s-phi-1}
\eea
for $m_{3/2} =10^{-4} - 10~{\rm GeV}$ and
\bea
|k_1| &\lesssim& O(10^{-9})   \,
\lrfp{m_\phi}{10^{12}{\rm\,GeV}}{-\frac{1}{2}} 
\lrfp{|\phi_0|}{10^{15}{\rm\,GeV}}{-1},\non\\
|k_2| &\lesssim& O(0.1)   \,
\lrfp{M_{\rm mess}}{10^{10}{\rm\,GeV}}{-1}
\lrfp{m_\phi}{10^{12}{\rm\,GeV}}{-\frac{1}{2}} 
\lrfp{|\phi_0|}{10^{15}{\rm\,GeV}}{-1},
\label{k-s-phi-2}
\eea
for $1$~keV $ \lesssim m_{3/2} \lesssim 10^{-4}$~GeV.  Thus 
$k_1$ is severely constrained as before, which indicates that
$k_1$ must be vanishing due to a symmetry.
The constraint on $k_2$  depends on the messenger scale $M_{\rm mess} = |y_M| M_s$; 
 it becomes severer for larger $M_{\rm mess}$.  Note that
the constraint is more or less similar to that obtained from the $\tilde \Phi$ decay
into the SM (s)particles [cf. (\ref{g-s-phi-1}) and
(\ref{g-s-phi-2})].

Lastly we comment on the lightest messenger particle (LMP). If the
inflaton dominantly decays into the messengers, the LMP is also
produced.  Note that, since the messenger number is conserved in
Eq.~(\ref{eq:gmsb}) and in the gauge interactions, all the produced
$\psi_\Psi$ and $\psi_{\bar\Psi}$ eventually decay into the lightest
messenger particle, which is a combination of the bosonic components
of the messenger fields.  If $T_R$ exceeds $M_{\rm\,mess}$, the LMPs
are thermalized, while, if not, they are non-thermally produced.  It
has been known that thus generated LMP easily overcloses the universe
if it is stable~\cite{Dimopoulos:1996gy}. So it must be unstable due
to a direct or indirect interaction with the visible sector.  If the
LMP decays fast enough, the constraints (\ref{k-s-phi-1}) and
(\ref{k-s-phi-2}) are valid. However, if the LMP decay rate is small
enough, they may dominate the universe and produce large entropy at
late time, diluting the pre-existing
gravitinos~\cite{Baltz:2001rq+Fujii:2002fv+Jedamzik:2005ir}.  In this
case the constraints (\ref{k-s-phi-1}) and (\ref{k-s-phi-2}) cannot be
applied.  The detailed discussion on the LMP abundance and its effect
on the thermal history may be important for constraining the mixings,
but it is beyond the scope of this paper.

\section{Other Issues}
\label{sec:5}

So far, we have assumed that it is the heavy scalar field
$\tilde{\Phi}$ which dominates the energy density of the universe and
which causes the reheating. However, there is a potential cosmological
problem of the $z$ modulus field, which gives dominant contribution to
the SUSY breaking, i.e., $G^z G_z\simeq 3$. In fact, as we have seen
in Sec.~\ref{sec:2}, its mass is comparable to the gravitino mass
unless there is a significant SUSY breaking effect on the $z$ mass. 
Since the $z$ field, which
corresponds to $\tilde Z$ in the mass eigenstate basis, couples with
the visible sector only via the non-renormalizable interactions, it
may cause the moduli problem of itself. The importance of the $\tilde Z$
field for cosmology was also mentioned in Ref.~\cite{Lebedev:2006qq}.

The evolution of the energy density of $\tilde Z$ field depends on the
model and cosmological scenario. In fact $\tilde Z$ might be displaced
far from the potential minimum during the inflation, which would lead
to an universe dominated by the $\tilde Z$'s oscillation. 
To be concrete, let us consider the gravity mediation.
Then, if $m_z\gsim m_{3/2}$, the $\tilde Z$ decay would produce too
many gravitinos with $B_{3/2}\simeq O(1)$. In addition, $\tilde Z$
must decay before the BBN starts. Therefore the $\tilde Z$--dominated
universe could be consistent only if $m_{3/2}\gsim m_z\gsim 100$ TeV,
which is a very challenging constraint on the structure of the SUSY
breaking sector. Note that, even if
the initial displacement of the $z$ field is set to be zero in the
model basis by some mechanism, the mass eigenstate $\tilde Z$ can
obtain a finite amplitude after $\phi$ starts oscillating, through the
$\phi-z$ mixing.  Since the thermal history associated with the decay
of the SUSY breaking field is strongly dependent on the detailed
structure of the SUSY breaking sector as well as the cosmological
scenario, further studies are required for this issue.

In the low energy SUSY breaking models, $\tilde Z$ may have 
stronger couplings with the messenger and/or the visible sector,
through which $\tilde Z$ decays fast enough. There is an additional field
scalar $s$ in the messenger sector, which may cause a similar
problem. However, the potential $s$-modulus
problem is not serious if (\ref{eq:ineq}) is satisfied.

So far we have discussed the SUSY breaking models that contain
direct couplings between the visible sector and the SUSY breaking field.
In the case of the anomaly mediation~\cite{AM}, the visible sector is
sequestered from the SUSY breaking sector, for example, by the
geometrical separation.  Since the sequestered K\"ahler potential is
not minimal, the models generally contain finite mixings. Then the
gravitino and $\tilde Z$ productions can be one of the dominant
channels of the $\tilde\Phi$ decay. The distinct difference from the
gravity mediation lies in the interactions between the SUSY breaking
field and the visible sector: they are generally quite suppressed
because of the sequestering. Thus, in the anomaly mediation, we need
to investigate the subsequent decay of the SUSY breaking field with a
special care, and to this end, we must go into details of the hidden
sector.  For instance, the minimal setup of the anomaly mediation is
known to suffer from the tachyonic sleptons. To cure the
charge-breaking vacuum, one might introduce an extra field to mediate
the SUSY breaking effects. Then the field may affect the cosmological
scenario related to the $\tilde\Phi$ decay as well as that of $\tilde
Z$. Thus, the analysis quite depends on the models.

\section{Conclusions and discussion}
\label{sec:6}

In this paper, we studied the decay processes of the heavy scalar
$\phi$, especially paying attention to the effects of the mixture
between $\phi$ and the SUSY breaking field, $z$. The scalar field
generally mixes with the SUSY breaking field in the K\"ahler
potential. Then the decay amplitudes of the heavy scalar field into
the lighter particles can be modified significantly by the
mixing-induced interactions. We explicitly estimated the production
rates of the SM (s)particles, gravitino and the SUSY breaking
field. In particular, we obtained the general form of the gravitino
coupling in the mass-eigenstate basis.

The mixture of $\phi$ with the SUSY breaking field is particularly
important for the thermal history, once the field $\phi$ dominates the
energy density of the universe. Such a scalar field may be identified,
for example, with a modulus and inflaton. In this paper, we also
discussed the impacts on the cosmology due to the mixing-induced decay
both in the modulus and inflaton cases. Particularly, it was found
that the modulus decay generally suffers from the moduli-induced
gravitino problem. In the inflaton decay, it is stressed that the
mixture, if any, provides a lower bound on the reheating temperature
because the inflaton can decay into the SM (s)particles via the
interaction that mediates the SUSY breaking effects to the visible
sector.  In the GMSB setup, the inflaton may rapidly decay into the
SM (s)particles or the messengers due to the $\phi - s$ mixing, resulting in too high
reheating temperature.  Such a feature becomes prominent for the
models with a large inflaton mass and VEV, like the hybrid inflation
model. As well as the gravitino overproduction problem due to the
mixings in the K\"ahler potential, such a high temperature suffers
from too much abundance of the gravitino produced by the thermal
scatterings.

All these difficulties are originated from the mixture between the
heavy scalar and the SUSY breaking sector fields. One of the simplest
solutions, especially for the inflaton, is to postulate a symmetry of
$\phi$ which is preserved at the vacuum. In many inflation models, the
inflaton acquires a VEV in the vacuum, therefore the mixings are not
protected by any symmetries.  In a simple class of the chaotic
inflation, however, the inflaton field is invariant under a $Z_2$
discrete symmetry, $\phi \rightarrow -\phi$. Then the scalar VEV as
well as the auxiliary component of the inflaton will be
vanishing. Thus the inflaton field does not mix with the SUSY breaking
field in this case.  Another solution is to introduce large entropy
production at a late time.  However, we always need to pay attention
to whether the additional field that induces the entropy dilution is
free from the mixing with the SUSY breaking sector field or not.

It is a symmetry that determines whether a field mixes with another,
since the symmetry dictates structure of the interactions. Once the
symmetry is broken spontaneously or explicitly, there is no reason
that the mixings should not occur.  To construct a successful
cosmological scenario, one must always check whether the mixings might
affect the dynamics concerned.  Although this might involve the
detailed structure of e.g. the SUSY breaking sector, we believe that
thus obtained constraints on the mixings will shed light on the true
structure of the high energy physics.


\section*{acknowledgement}

We thank M.~Kawasaki and T.~T.~Yanagida for discussions, and
M.~Ibe and Y.~Shinbara for useful comments.
M.E. and F.T.  would like to thank the Japan Society for Promotion of Science
for financial support.

\appendix

\section{Decay into the gravitinos in the goldstino picture}
\label{appa}
According to the goldstino--equivalence
theorem~\cite{Casalbuoni:1988kv}, the scalar--gravitino--gravitino
interaction discussed in Refs.~\cite{Endo:2006zj,Nakamura:2006uc}
should also be understood in the goldstino limit, i.e., in the context
of spontaneously broken global SUSY. Here we show it explicitly. The
generic form of the scalar--goldstino--goldstino interaction has been
derived in Ref.~\cite{Brignole:2003cm} in the context of
Higgs--goldstino--goldstino interaction:
\begin{eqnarray}
  \mathcal{L} = \frac{1}{2F_{\rm total}}
  \left(\frac{\vev{F^i}}{F_{\rm total}}\right)
  \left(M^2_{ij^*} \varphi^{*j} + M^2_{ij} \varphi^j \right)
  \overline{\widetilde{\chi}} P_L \widetilde{\chi}
  \;+\; \mathrm{h.c.},
\end{eqnarray}
where $M^2_{ij^*}=\vev{V_{ij^*}}$, $M^2_{ij}=\vev{V_{ij}}$ and
$\widetilde{\chi}$ is the (4-component) goldstino field. Notice that,
when interpreted in terms of SUGRA, the interaction {\it does} have an
enhancement factor $1/F_{\rm total}\propto 1/m_{3/2}$, and there is no
chirality suppression. Assuming $M^2_{ij^*} \gg M^2_{ij}$ and taking a
basis where $M^2_{ij^*} = \delta_{ij} m_i^2$, we obtain
\begin{eqnarray}
  \mathcal{L}_{\phi\tilde{\chi}\tilde{\chi}} = \frac{m_\phi^2}{2F_{\rm total}}
  \left(\frac{\vev{F_\phi}}{F_{\rm total}}\right)
  \phi^*
  \overline{\widetilde{\chi}} P_L \widetilde{\chi}
  \;+\; \mathrm{h.c.},
\end{eqnarray}
leading to
\begin{eqnarray}
  \Gamma(\phi_{\rm R,I}\to \widetilde{\chi}\widetilde{\chi})
  =
  \frac{1}{32\pi}\frac{m_\phi^5}{F_{\rm total}^2}
  \left(\frac{|\vev{F_\phi}|}{F_{\rm total}}\right)^2\;.
\end{eqnarray}
This can be rewritten in terms of SUGRA by using $F_{\rm total} =
\sqrt{3}m_{3/2}M_P$:
\begin{eqnarray}
  \Gamma(\phi_{\rm R,I}\to 2\psi_{3/2})
  =
  \frac{1}{96\pi}\frac{m_\phi^5}{m_{3/2}^2 M_P^2}
  \left(\frac{|\vev{F_\phi}|}{F_{\rm total}}\right)^2\;.
\end{eqnarray}
which, by using $F_\phi/F_{\rm total}=G_\phi/\sqrt{3}$,
reproduces the result obtained in Refs.~\cite{Endo:2006zj,Nakamura:2006uc}. 
Whether it is suppressed or enhanced by the gravitino mass depends on the 
fractional contribution of the $\phi$--multiplet to the total amount of the 
SUSY breaking, $\vev{F_\phi}/F_{\rm total}$. In the extreme case where $\phi$ 
itself is the dominant source of the SUSY breaking, $\vev{F_\phi}/F_{\rm total} 
\simeq 1$, the rate is indeed enhanced. For $\vev{F_\phi}/F_{\rm total} \simeq 
m_{3/2}/m_\phi$ the $m_{3/2}$ dependence cancels out, and for $\vev{F_\phi} 
/ F_{\rm total}\lsim (m_{3/2}/m_\phi)^2$, the rate is suppressed by the gravitino 
mass.


\begin{thebibliography}{99}

\bibitem{Endo:2006zj}
  M.~Endo, K.~Hamaguchi and F.~Takahashi,
  arXiv:hep-ph/0602061.

\bibitem{Nakamura:2006uc}
  S.~Nakamura and M.~Yamaguchi,
  arXiv:hep-ph/0602081.

\bibitem{Kawasaki:2006gs}
  M.~Kawasaki, F.~Takahashi and T.~T.~Yanagida,
  arXiv:hep-ph/0603265.

\bibitem{Asaka:2006bv}
  T.~Asaka, S.~Nakamura and M.~Yamaguchi,
  arXiv:hep-ph/0604132.

\bibitem{Dine:2006ii}
  M.~Dine, R.~Kitano, A.~Morisse and Y.~Shirman,
  arXiv:hep-ph/0604140.
  
  \bibitem{GMSB}
  M.~Dine, A.~E.~Nelson and Y.~Shirman,
  Phys.\ Rev.\ D {\bf 51} (1995) 1362;
  M.~Dine, A.~E.~Nelson, Y.~Nir and Y.~Shirman,
  Phys.\ Rev.\ D {\bf 53} (1996) 2658;
  For a review, see, for example, 
  G.~F.~Giudice and R.~Rattazzi,
  Phys.\ Rep.\  {\bf 322} (1999) 419,
  and references therein.
  
\bibitem{Lebedev:2006qq}
O.~Lebedev, H.~P.~Nilles and M.~Ratz,
Phys.\ Lett.\ B {\bf 636}, 126 (2006).

\bibitem{Buchmuller:2004rq}
W.~Buchmuller, K.~Hamaguchi, M.~Ratz and T.~Yanagida,
Phys.\ Lett.\ B {\bf 588}, 90 (2004).

\bibitem{Hashimoto:1998mu}
M.~Hashimoto, K.~I.~Izawa, M.~Yamaguchi and T.~Yanagida,
Prog.\ Theor.\ Phys.\  {\bf 100}, 395 (1998).

\bibitem{KYY}
  K.~Kohri, M.~Yamaguchi and J.~Yokoyama,
  Phys.\ Rev.\ D {\bf 72}  (2005) 083510;
  Phys.\ Rev.\ D {\bf 70} (2004) 043522.
  
\bibitem{WessBagger}
  J. Wess and J. Bagger, Supersymmetry and Supergravity,
  (Princeton Unversity Press, 1992).  

  
   \bibitem{Kawasaki:1999na}
  M.~Kawasaki, K.~Kohri and N.~Sugiyama,
  Phys.\ Rev.\ Lett.\  {\bf 82}, 4168 (1999);
  Phys.\ Rev.\ D {\bf 62}, 023506 (2000);
  K.~Ichikawa, M.~Kawasaki and F.~Takahashi,
  Phys.\ Rev.\ D {\bf 72}, 043522 (2005).
  
\bibitem{Hannestad:2004px}
S.~Hannestad,
Phys.\ Rev.\ D {\bf 70}, 043506 (2004).

\bibitem{Kawasaki:2004yh}
  M.~Kawasaki, K.~Kohri and T.~Moroi,
  Phys.\ Lett.\ B {\bf 625}, 7 (2005);
  Phys.\ Rev.\ D {\bf 71}, 083502 (2005).

\bibitem{Kohri:2005wn}
K.~Kohri, T.~Moroi and A.~Yotsuyanagi,
arXiv:hep-ph/0507245.

\bibitem{Kumekawa:1994gx}
  K.~Kumekawa, T.~Moroi and T.~Yanagida,
  Prog.\ Theor.\ Phys.\  {\bf 92}, 437 (1994).
\bibitem{Izawa:1996dv}
  K.~I.~Izawa and T.~Yanagida,
  Phys.\ Lett.\ B {\bf 393}, 331 (1997).

\bibitem{Copeland:1994vg}
  E.~J.~Copeland, A.~R.~Liddle, D.~H.~Lyth, E.~D.~Stewart and D.~Wands,
  Phys.\ Rev.\ D {\bf 49}, 6410 (1994).
\bibitem{Dvali:1994ms}
  G.~R.~Dvali, Q.~Shafi and R.~K.~Schaefer,
  Phys.\ Rev.\ Lett.\  {\bf 73}, 1886 (1994).
\bibitem{Linde:1997sj}
  A.~D.~Linde and A.~Riotto,
  Phys.\ Rev.\ D {\bf 56}, 1841 (1997).


\bibitem{Moroi:1993mb}
  T.~Moroi, H.~Murayama and M.~Yamaguchi,
  Phys.\ Lett.\ B {\bf 303}, 289 (1993).
  
\bibitem{Bolz:2000fu}
M.~Bolz, A.~Brandenburg and W.~Buchmuller,
Nucl.\ Phys.\ B {\bf 606}, 518 (2001).

\bibitem{Dimopoulos:1996gy}
S.~Dimopoulos, G.~F.~Giudice and A.~Pomarol,
Phys.\ Lett.\ B {\bf 389}, 37 (1996).

\bibitem{Baltz:2001rq+Fujii:2002fv+Jedamzik:2005ir}

   E.~A.~Baltz and H.~Murayama,
   JHEP {\bf 0305}, 067 (2003).

M.~Fujii and T.~Yanagida,
Phys.\ Lett.\ B {\bf 549}, 273 (2002).

K.~Jedamzik, M.~Lemoine and G.~Moultaka,
Phys.\ Rev.\ D {\bf 73}, 043514 (2006).



\bibitem{AM}
  L.~Randall and R.~Sundrum,
  Nucl.~Phys. {\bf B557}, 79  (1999);
  G.F.~Giudice, M.A.~Luty, H.~Murayama and R.~Rattazzi,
  JHEP {\bf 9812}, 027 (1998);
  J.A.~Bagger, T.~Moroi and E.~Poppitz,
  JHEP {\bf 0004}, 009 (2000).

\bibitem{Casalbuoni:1988kv}
  R.~Casalbuoni, S.~De Curtis, D.~Dominici, F.~Feruglio and R.~Gatto,
  Phys.\ Lett.\ B {\bf 215}, 313 (1988).

\bibitem{Brignole:2003cm}
  A.~Brignole, J.~A.~Casas, J.~R.~Espinosa and I.~Navarro,
  Nucl.\ Phys.\ B {\bf 666}, 105 (2003).


\end{thebibliography}
\end{document}